# First-principles study of elastic properties of Cr-Al-N

Liangcai Zhou,[1,*] David Holec,[2] and Paul. H. Mayrhofer[1]

[1]*Institute of Materials Science and Technology, Vienna University of Technology, A-1040 Vienna, Austria*

[2]*Department of Physical Metallurgy and Materials Testing, Montanuniversität Leoben, A-8700 Leoben, Austria*

The elastic properties of paramagnetic cubic B1 (c-) $Cr_{1-x}Al_xN$ ternary alloys are studied using stress-strain and energy-strain methods within the framework of Density Functional Theory (DFT). A strong compositional dependence of the elastic properties is predicted. Young's modulus, $E$, and shear modulus, $G$, exhibit the same compositional trends as experimentally measured hardness values (i.e. increasing with Al content), while bulk modulus, $B$, remains almost constant. The isotropic elastic response in the c-$Cr_{1-x}Al_xN$ is predicted for concentrations around $x$=0.50. Brittle behavior and directional bonding characteristics are predominant in the c-$Cr_{1-x}Al_xN$ coatings in the whole composition range, and become more pronounced with increasing Al content.

[*]Author to whom correspondence should be addressed;

E-mail address: zhlc1985@gmail.com, david.holec@unileoben.ac.at

I. INTRODUCTION

$Cr_{1-x}Al_xN$ thin films are nowadays commonly utilized as protective coatings in machining, automobile and other industrial areas, due to their outstanding mechanical, chemical, electrical, and thermal properties.[1,2] These beneficial effects are obtained for the cubic B1 structure (NaCl prototype). The main advantages of $Cr_{1-x}Al_xN$ in comparison with $Ti_{1-x}Al_xN$ are that (i) $Cr_{1-x}Al_xN$ solves more Al in the cubic phase, and (ii) the stability of c-$Cr_{1-x}Al_xN$ coatings is preserved up to higher temperatures in oxidative environments, as both chromium and aluminum form protective dense oxides with reduced diffusivity for the species involved.[3-6] Therefore a substantial amount of experimental results have been published concentrating on the synthesis, structure and properties of c-$Cr_{1-x}Al_xN$ and their alloys.[4, 7-12]

The stoichiometric compound CrN has a rock-salt (NaCl prototype) paramagnet (PM) cubic structure at room temperature. Below the Néel temperature ($T_N$=273-283K), it is antiferromagnetic (AFM) with a distorted orthorhombic structure, showing a small structural distortion from the underlying B1 lattice characterized by the angle α≈88.3°.[13, 14] The AFM order consists of alternating double (110)-planes of Cr atoms with spin up and spin down, respectively. The transition from cubic to orthorhombic (*Pnma*) is a first-order phase transition and results in a discontinuous volume reduction by ~0.59%.[15, 16] A number of theoretical studies taking different magnetic states into account, including AFM, ferromagnetic (FM), PM and non-magnetic (NM) states proved that CrN above the room temperature should be treated with PM state.[16-20]

The theoretically predicted maximum solubility of AlN in c-$Cr_{1-x}Al_xN$ varies from 0.48 to 0.95, depending on different structural configurations, external condition such as pressure, or used methods.[21-25] For higher Al content, the preferred structure is wurtzite B4 (ZnS prototype). Mixing enthalpy of c-$Cr_{1-x}Al_xN$ shows that c-$Cr_{1-x}Al_xN$ is more stable against decomposition than c-Ti$_{1-x}$Al$_x$N.[26, 27] Tasnádi *et al.* showed that the elastic properties of Ti$_{1-x}$Al$_x$N are strongly affected by the Al content.[28] The metal sublattice population was shown to influence also the elastic properties of c-Cr$_{1-x}$Al$_x$N,[22, 27] but a thorough study is still missing. Information on the elastic properties is important for understanding and predicting the mechanical and dynamical behavior, and for indications of material-related properties such as hardness, brittleness, ductility, fracture toughness, and bond characteristics necessary for a knowledge-based coatings design.

The aim of the present paper is to use first-principles calculations to theoretically investigate the compositional dependence of the elastic properties of the c-Cr$_{1-x}$Al$_x$N system. The special quasi-random structures (SQS)[29] approach is applied to model both, the chemical as well as magnetic disorder representing the paramagnetic state of the (Cr$_{0.5}^{\uparrow}$Cr$_{0.5}^{\downarrow}$)$_{1-x}$Al$_x$N system. We first obtained structural properties (lattice parameters, energies of formation) to compare our results with those from literature. Further, we calculated the single crystal elastic properties using two methods, the energy-strain and stress-strain method. Subsequently, we applied the self-consistent Hershey homogenization scheme[30] to obtain the isotropic polycrystalline estimates of the elastic response.

## II. METHODOLOGY

Mixing of Cr$^{\uparrow}$, Cr$^{\downarrow}$ and Al atoms takes place on one sublattice following the SQS methodology, while the other sublattice is fully occupied with N atoms. 3x3x2 (36 atoms) and 2x2x2 (32 atoms) supercells are used for the cubic B1 and wurtzite B4 structures, respectively. The short range order parameters (SROs) are optimized for pairs at least up to fifth order. The thus generated structures together with the corresponding SROs are listed in Appendix (Table II to Table IV).

The density functional theory based calculations are performed using the Vienna Ab initio Simulation Package (VASP).[31, 32] The ion-electron interactions are described by the projector augmented wave method (PAW)[33] with an energy cutoff of 500 eV, and the generalized gradient approximation (GGA) for the exchange-correction effects, which is parameterized by Perdew–Burke–Ernzerhof (PBE).[34] The $k$-point meshes are 6×6×9 and 7×7×5 for the cubic B1 and the wurtzite B4 structures, respectively. The energy convergence criterion for electronic self-consistency is 0.1 meV/atom.

In principle, there are two routs of computing single crystal elastic constants from first-principles calculations: the energy-strain approach and the stress-strain approach.[35, 36] The energy-strain approach[36] is based on the computed total energies of properly selected strained states of the crystal. The second order elastic constants appear, via Hooke's law, in the second order Taylor expansion coefficient of the strain energy. Assuming the cubic symmetry of the cubic supercell, three independent elastic constants, $C_{11}$, $C_{12}$ and $C_{44}$, are to be determined. Two deformation modes with keeping the unit cell volume constant[37] are employed and the bulk modulus, $B$, is derived from the Birch-Murnaghan equation of state (EOS).[38] More details about the derivation of the elastic constants of cubic B1 structure can be found in Ref. 37.

The stress-strain approach, on the other hand, utilizes the stress tensor calculated by VASP[31, 32]. Hence the elastic constants can be directly derived from the generalized Hooke's law. A set of strains $\boldsymbol{\varepsilon} = (\varepsilon_1, \varepsilon_2, \varepsilon_3, \varepsilon_4, \varepsilon_5, \varepsilon_6)$ (where $\varepsilon_1$, $\varepsilon_2$ and $\varepsilon_3$ are the normal strains, $\varepsilon_4$, $\varepsilon_5$ and $\varepsilon_6$ are the shear strains in Voigt's notation[39]) is imposed on a crystal, by alternating the unit cell lattice vectors ($\overline{R}$) from the original ($R$) as follows:

$$\overline{R} = R \begin{pmatrix} 1+\varepsilon_1 & \varepsilon_6/2 & \varepsilon_5/2 \\ \varepsilon_6/2 & 1+\varepsilon_2 & \varepsilon_4/2 \\ \varepsilon_5/2 & \varepsilon_4/2 & 1+\varepsilon_3 \end{pmatrix} \quad (1)$$

A 6×6 elastic constants matrix, **C**, with components of $C_{ij}$ in Voigt's notation, relates the strain vector **ε** with the stress vector **σ** =($\sigma_1$, $\sigma_2$, $\sigma_3$, $\sigma_4$, $\sigma_5$, $\sigma_6$) as **σ** = **ε C**. In the present work, the following six linearly independent sets of strains are applied:

$$\begin{pmatrix} x & 0 & 0 & 0 & 0 & 0 \\ 0 & x & 0 & 0 & 0 & 0 \\ 0 & 0 & x & 0 & 0 & 0 \\ 0 & 0 & 0 & x & 0 & 0 \\ 0 & 0 & 0 & 0 & x & 0 \\ 0 & 0 & 0 & 0 & 0 & x \end{pmatrix} \quad (2)$$

where each row is one set of strains **ε** with $x$ being a normal (or shear) strain $\varepsilon_i$. After $n$ first-principles calculations upon the deformed lattices due to $n$ sets of strains **ε**, the corresponding stress matrix **σ** with $n$ sets of stresses is obtained. Based on Hook's law, the elastic stiffness constants matrix, **C**, is determined as,

$$\mathbf{C} = \mathbf{\varepsilon}^{-1} \mathbf{\sigma} \quad (3)$$

where '-1' represents the pseudo-inverse, solved by the singular value decomposition method.

Recently, Tasnádi et al. pointed out that using SQS approach to model the disordered state results in 21 elastic constant elements in the elastic tensor matrix of disordered solid solution, due to the point group symmetry broken by the SQS approach.[40] This implies the need for applying 21 independent deformation modes for deriving 21 elastic constants using the energy-strain approach. Such a treatment is extremely computationally expensive. The stress-strain approach, on the contrary, is much more efficient as the six strains (Eq. 2) are sufficient to obtain a full elastic constants matrix.[35, 41] Finally, the macroscopic cubic elastic constant, $\overline{C_{11}}$, $\overline{C_{12}}$, and $\overline{C_{44}}$, are obtained by simple averaging:[42]

$$\overline{C_{11}} = \frac{C_{11} + C_{22} + C_{33}}{3} \qquad (4)$$

$$\overline{C_{12}} = \frac{C_{12} + C_{13} + C_{23}}{3} \qquad (5)$$

$$\overline{C_{44}} = \frac{C_{44} + C_{55} + C_{66}}{3} \qquad (6)$$

Calculations are performed from $x=\pm 0.007$ up to $\pm 0.042$ with a step equal to 0.007, and different $k$-point meshes 4×4×6 to 12×12×18, indicating that the predicted errors of $C_{ij}$ are quite small (< 2%). Finally, $x=\pm 0.007$ and $k$-point sampling 6×6×9 are chosen in the present work. Based on the single crystal elastic constants $C_{ij}$, the isotropic equivalents of polycrystalline properties are computed via the self-consistent Hershey approach.[30]

## III. RESULTS AND DISCUSSION

### A. Equilibrium properties

We performed the calculations of c-$Cr_{1-x}Al_xN$ in the PM and NM states and w-$Cr_{1-x}Al_xN$ in the PM state. The calculated energy of formation, $E_f$, as a function of the AlN content, $x$, is presented in Fig. 1, where third-order polynomial fittings serve as guidelines for the eye. The $E_f$ differences of c-$Cr_{1-x}Al_xN$ in the PM and NM states are significant; $E_f$ differences of PM c-$Cr_{1-x}Al_xN$ is up to 0.1 eV/atom lower than that of NM state. This suggests that the PM state cannot be neglected. Comparing $E_f$ of c- and w-$Cr_{1-x}Al_xN$ in the PM state yields the maximum solubility of AlN in c-Cr$_{1-x}$Al$_x$N at around x≈0.75, a value consistent with previous experimental and theoretical calculation results.[21, 22]

The calculated lattice parameter, *a*, and bulk modulus, *B*, of c-Cr$_{1-x}$Al$_x$N in the PM and NM states as a function of Al content, *x*, are shown in Figs. 2a and 2b. As demonstrated in Fig. 2a, the calculated lattice parameters of c-Cr$_{1-x}$Al$_x$N in the PM and NM states as functions of AlN content have different trends. The lattice parameter of PM c-Cr$_{1-x}$Al$_x$N decreases from 4.145 Å to 4.069 Å as AlN content increases from 0 to 1, agreeing well with those previously reported in the literature[22], which is well supported by experimental results.[4, 7, 43, 44] Contrarily, the lattice parameters of NM c-Cr$_{1-x}$Al$_x$N exhibit a small increase as AlN content increases from 0 to 1. Similar to other Al-containing cubic transition metal nitrides, also here we predict a positive bowing away from the Vegard's-like behavior.[45] The respective bowing parameters are 0.034Å and 0.024Å for PM and NM states, respectively. The inset in Fig. 2a shows the positive bowing of lattice parameters from Vegard's rule for the PM and NM states. The bowing is attributed to a gradual change of the bonding character with Al content. The bulk modulus, *B*, of c-Cr$_{1-x}$Al$_x$N in the PM state presented in Fig. 2b is almost constant as increases from 0 to 1 (overall change of ≈5 GPa), while NM state calculations for c-Cr$_{1-x}$Al$_x$N show a steep decreasing trend, with *B* decreasing from 329 to 252 GPa as increases from 0 to 1. The bulk modulus *B* of c-Cr$_{1-x}$Al$_x$N in the PM state is smaller than that of the NM state in the whole composition range. This is mostly the effect of different equilibrium volumes of those two phases: according to the Murnaghan equations of state,[38] the bulk modulus *B* increases with the pressure, $B = B_0 + \frac{dB_0}{dP}P$, i.e., with decreasing the lattice parameter at certain composition. Indeed, when e.g., the PM bulk modulus is extrapolated to the NM equilibrium volume, a value close to that of NM $B_0$ is obtained. In agreement with Alling *et al.*,[17] we once again conclude that the proper consideration of the PM state is essential for the $Cr_{1-x}Al_xN$ alloy.

B. Elastic properties

The compositional trends in the elastic response of c-$Cr_{1-x}Al_xN$ in the PM state are studied using energy-strain and stress-strain methods. All independent components of the c-$Cr_{1-x}Al_xN$, as calculated by the stress-strain method, are summarized in table I. The thus obtained elastic tensors exhibit small deviations from a strict cubic symmetry, similarly to the results of Tasnádi *et al*[40] for $Ti_{0.5}Al_{0.5}N$. Figure 3a presents the obtained *ab initio* predicted cubic single-crystal elastic constants, $C_{11}$, $C_{12}$ and $C_{44}$, of c-$Cr_{1-x}Al_xN$ as a function of Al content.[46] For all compositions, the elastic constants fulfill the Born stability criteria[39] for cubic crystals,

$$C_{44}>0,\ C_{11}>|C_{12}|,\ C_{11}+2\ C_{12}>0 \qquad (7)$$

The elastic constants from stress-strain and energy-strain methods show a good agreement. The elastic constants $C_{11}$ of c-$Cr_{1-x}Al_xN$ are significantly stiffer than the other two elastic constants, $C_{12}$ and $C_{44}$, in particular for the Cr-rich composition. $C_{12}$ and $C_{44}$ constants increase with the amount of Al, while $C_{11}$, in contrast, shows a small decrease as Al content increases. These compositional trends in the $Cr_{1-x}Al_xN$ system are very similar to those predicted for the $Ti_{1-x}Al_xN$ system.[28] Some insight in these trends brings the analysis of the chemical bonding. A strong hybridization of $sp^3d^2$ orbitals accompanied with a weak hybridization of second next-nearest neighboring Cr *d* state takes place in CrN. AlN, on the other hand, is a semiconductor with bonding of an ionic-covalent nature due to significantly larger charger transfer from cation to anion accompanying a strong Al 3*s*-N 2*p* hybridization.[47-50] We can therefore conclude that the weak *d-d* bonding (metallic interaction) in CrN is responsible for the low value of the $C_{44}$ elastic constant, while its high value in AlN is caused by strong Al-Al and N-N repulsion as they get closer under the shear deformation.

The calculated polycrystalline estimates of Young's modulus, $E$, and shear modulus, $G$, are given in Fig. 3b together with bulk modulus, $B$. The results from stress-strain and energy-strain methods agree well, which is not surprising given the excellent agreement of the single-crystal elastic constants. Values of the bulk modulus, $B$, obtained by the strain-energy method are those derived from the Birch-Murnaghan EOS i.e., those used to obtain the single crystal elastic constants by this method. On the contrary, the strain-stress method allowed us to obtain the single crystal elastic constants without using $B$ form the EOS, and thus to independent estimate for $B$. The fact that the two datasets for $B$ almost overlap is a proof of the consistency of our calculations. The Young's modulus, $E$, and shear modulus, $G$, change smoothly and increase with the amount of Al. Both of them show almost a linear dependence on the composition in contrast to an almost practically constant value of $B$. By analyzing the hardness of cubic metal mononitrides, Fulcher *et al*. have recently pointed out that the Young's and shear moduli exhibit better correlations with hardness than bulk modulus.[51] Our results confirm this finding also for the case of the $Cr_{1-x}Al_xN$ alloy. Indeed, the majority of experimental results state an indentation hardness increase with Al-content for $Cr_{1-x}Al_xN$ from ~21 GPa[52] for CrN to ~32 GPa for c-$Cr_{1-x}Al_xN$ with Al contents close to the B1/B4 transition.[12]

To quantify the elastic anisotropy in the c-$Cr_{1-x}Al_xN$ system, the Zener's anisotropy ratio,[53]

$$A=2C_{44}/(C_{11}-C_{12}) \qquad (8)$$

together with the ratio between the directional Young's modulus $E_{<111>}$ and $E_{<100>}$ are presented in Fig. 4a. The results indicate that the stiffest direction in c-$Cr_{1-x}Al_xN$ changes from <100> to

<111> direction as Al content increases. This is caused by softening of $C_{11}$, (weakening of the $sp^3d^2$ hybridization) and stiffening of $C_{44}$ and $C_{12}$ elastic constants (increasing charge transfer) with the addition of Al. It is also worth noting that the ratio $A$ and $E_{<111>}/E_{<100>}$ are close to 1 at certain concentration. Herein highlighted in Fig. 4a for the stress-strain method, the isotropic composition is around $x=0.44$, while it equals to approximately 0.56 for the energy-strain method. We therefore predict that the directional dependence of Young's modulus almost diminishes for alloys with compositions close to $Cr_{0.5}Al_{0.5}N$.

Figure 4b shows Poisson's ratio, $v$, and the $B/G$ ratio as a function of Al content, $x$. Frantsevich *et al*. suggested that the Poisson's ratio can be used as an indicator for ductile ($v>1/3$) or brittle behavior ($v<1/3$).[54] According to this criterion, c-$Cr_{1-x}Al_xN$ can be regarded as a brittle material, since $v$ is smaller than 1/3 over the whole composition range. Another criterion for ductility or brittleness is the value of the $B/G$ ratio. The higher or lower the $B/G$ ratio is, the more ductile or brittle the material is, respectively. The critical value which separates ductile and brittle materials is approximately 1.75.[55] The $B/G$ ratios calculated from stress-strain and energy-strain methods are again consistent with each other, and are below or close to 1.75. Hence, the $B/G$ ratio in agreement with the Poisson's ratio criterion, confirms the brittle properties of c-$Cr_{1-x}Al_xN$ over the whole composition range. The brittleness increases with increasing Al content resulting from gradually weakening of *d-d* metallic bonding in the c-$Cr_{1-x}Al_xN$ with increasing Al content $x$.

It has been suggested in the literature that the Cauchy pressure,[56] $C_{12}$- $C_{44}$, can be used to characterize the bonding type. Negative Cauchy pressure corresponds to more directional bonding, while positive values indicate predominant metallic bonding. The calculated Cauchy pressure for c-$Cr_{1-x}Al_xN$ alloys is presented in the lower panel of Fig. 4b. Their values are close to zero or negative in the whole composition range, indicating that the predominant bonding behavior in c-$Cr_{1-x}Al_xN$ alloys is directional. The decreasing Cauchy pressure with increasing Al content indicates a tendency towards stronger directional character of the bonds, which results in an increased resistance against shearing, a result reflected also by the increasing trend in $C_{44}$. As explained above, this is a consequence of the gradual change of the hybridization between Cr 3*d* and N 2*p* states to the ionic-covalent bonding between Al 3*p* and N 2*p* states. Indeed, Litimein *et al.*[48] pointed out that the Al 3*s*-N 2*p* hybridization in AlN is stronger than that of transitional metal nitrides due to the proximity of the N 2*p* and Al 3*s* orbital energies and also the shorter bond length of AlN, which also can be applied to CrN. In addition, the weak metal-metal *d-d* interaction in CrN is gradually reduced to zero as Al content increases. The stronger Al 3*s*-N 2*p* hybridization creates a significant resistance to initializing plastic flow by shear, thus resulting in an increased hardness consequently. Therefore, c-$Cr_{1-x}Al_xN$ coatings with high Al-content are intrinsically harder than the low Al containing alloys.

## IV. CONCLUSIONS

The equilibrium properties of c-$Cr_{1-x}Al_xN$ are calculated using first principles, indicating that the proper magnetic order (namely the paramagnetic state) cannot be neglected. The elastic properties of c-$Cr_{1-x}Al_xN$ are obtained by two independent methods, the stress-strain and the energy-strain methods, which show an excellent agreement. Using the Hershey model we obtained the compositionally dependent polycrystalline Young's and shear moduli. Both exhibit a strong correlation with hardness, while the bulk modulus $B$, being almost constant, seems to be a poor indicator for hardness. The isotropic elastic response of c-$Cr_{1-x}Al_xN$ is predicted for concentrations around $x$=0.50. Cubic structured $Cr_{1-x}Al_xN$ is a brittle material in the whole composition range, and its brittleness increases with the Al content. The Cauchy pressure suggests a tendency towards stronger directional character of the bonds as the Al content increases. All the strong compositional dependencies of the elastic response predicted for the c-$Cr_{1-x}Al_xN$ system are closely linked with the changes in the electronic structure and bonding nature. Based on the results obtained we can conclude that high Al containing c-$Cr_{1-x}Al_xN$ coatings are intrinsically harder than their low Al containing counterparts.

## ACKNOWLEDGMENTS

Financial support by the START Program (Y371) of the Austrian Science Fund (FWF) is greatly acknowledged. The authors are thankful to Prof. Jörg Neugebauer from The Max-Planck-Institut für Eisenforschung GmbH for stimulating discussion.

# References

[1] O. Knotek, F. Löffler, and H. J. Scholl, Surf. Coat. Technol. **45**, 53 (1991).

[2] R. Sanjinés, O. Banakh, C. Rojas, P. E. Schmid, and F. Lévy, Thin Solid Films **420-421**, 312 (2002).

[3] H. Willmann, P. H. Mayrhofer, P. O. Å. Persson, A. E. Reiter, L. Hultman, and C. Mitterer, Scripta Mater. **54**, 1847 (2006).

[4] P. H. Mayrhofer, H. Willmann, and A. E. Reiter, Surf. Coat. Technol. **202**, 4935 (2008).

[5] R. E. Galindo, J. L. Endrino, R. Martínez, and J. M. Albella, Spectrochim. Acta **65**, 950 (2010).

[6] M. Kawate, A. K. Hashimoto, and T. Suzuki, Surf. Coat. Technol. **165**, 163 (2003).

[7] M. Kawate, A. Kimura, and T. Suzuki, J. Vac. Sci. Technol., A **20**, 569 (2002).

[8] G. S. Kim and S. Y. Lee, Surf. Coat. Technol. **201**, 4361 (2006).

[9] P. H. Mayrhofer, H. Willmann, L. Hultman, and C. Mitterer, J. Phys. D: Appl. Phys. **41**, 155316 (2008).

[10] J. L. Mo and M. H. Zhu, Tribology International **41**, 1161 (2008).

[11] A. E. Reiter, V. H. Derflinger, B. Hanselmann, T. Bachmann, and B. Sartory, Surf. Coat. Technol. **200**, 2114 (2005).

[12] H. Willmann, P. H. Mayrhofer, L. Hultman, and C. Mitterer, J. Mater. Res. **23**, 2880 (2008).

[13] L. M. Corliss, N. Elliott, and J. M. Hastings, Phys. Rev. **117**, 929 (1960).

[14] J. D. Browne, P. R. Liddell, R. Street, and T. Mills, Phys. Status Solidi A **1**, 715 (1970).

[15] M. Nasr-Eddine and E. F. Bertaut, Solid State Commun. **9**, 717 (1971).

[16] A. Filippetti and N. A. Hill, Phys. Rev. Lett. **85**, 5166 (2000).

[17] B. Alling, T. Marten, and I. A. Abrikosov, Nat. Mater. **9**, 283 (2010).

[18] B. Alling, T. Marten, and I. A. Abrikosov, Phys. Rev. B **82** (2010).

[19] F. Rivadulla, et al., Nat. Mater. **8**, 947 (2009).

[20] A. Filippetti, W. E. Pickett, and B. M. Klein, Phys. Rev. B **59**, 7043 (1999).

[21] D. Holec, F. Rovere, P. H. Mayrhofer, and P. B. Barna, Scripta Mater. **62**, 349 (2010).

[22] P. H. Mayrhofer, D. Music, T. Reeswinkel, H. G. Fuß, and J. M. Schneider, Acta Mater. **56**, 2469

(2008).

[23] R. F. Zhang and S. Veprek, Acta Mater. **55**, 4615 (2007).

[24] P. Spencer, Zeitschrift für Metallkunde **92**, 1145 (2001).

[25] Y. Makino, ISIJ international **38**, 925 (1998).

[26] B. Alling, T. Marten, I. A. Abrikosov, and A. Karimi, J. Appl. Phys. **102**, 044314 (2007).

[27] F. Rovere, D. Music, S. Ershov, M. T. Baben, H. G. Fuss, P. H. Mayrhofer, and J. M. Schneider, J. Phys. D: Appl. Phys. **43** (2010).

[28] F. Tasnádi, I. A. Abrikosov, L. Rogstrom, J. Almer, M. P. Johansson, and M. Odén, Appl. Phys. Lett. **97**, 231902 (2010).

[29] S. H. Wei, L. G. Ferreira, J. E. Bernard, and A. Zunger, Phys. Rev. B **42**, 9622 (1990).

[30] A. V. Hershey, J. Appl. Mech **21**, 236 (1954).

[31] G. Kresse and J. Hafner, Phys. Rev. B **47**, 558 (1993).

[32] G. Kresse and J. Furthmüller, Phys. Rev. B **54**, 11169 (1996).

[33] G. Kresse and D. Joubert, Phys. Rev. B **59**, 1758 (1999).

[34] J. P. Perdew, K. Burke, and M. Ernzerhof, Phys. Rev. Lett. **77**, 3865 (1996).

[35] Y. Le Page and P. Saxe, Phys. Rev. B **65**, 104104 (2002).

[36] Y. Le Page and P. Saxe, Phys. Rev. B **63**, 174103 (2001).

[37] D. Holec, M. Friák, J. Neugebauer, and P. H. Mayrhofer, Phys. Rev. B **85**, 064101 (2012).

[38] F. Birch, Phys. Rev. **71**, 809 (1947).

[39] J. F. Nye, *Physical Properties of Crystals, Their Representation by Tensors and Matrices* (Oxford University Press, Oxford, 1985).

[40] F. Tasnádi, M. Odén, and I. A. Abrikosov, Phys. Rev. B **85**, 144112 (2012).

[41] S. Shang, Y. Wang, and Z. K. Liu, Appl. Phys. Lett. **90**, 101909 (2007).

[42] M. Moakher and A. N. Norris, J ELASTICITY. **85**, 215 (2006).

[43] A. Kimura, M. Kawate, H. Hasegawa, and T. Suzuki, Surf. Coat. Technol. **169**, 367 (2003).

[44] Y. Makino and K. Nogi, Surf. Coat. Technol. **98**, 1008 (1998).

[45] D. Holec, R. Rachbauer, L. Chen, L. Wang, D. Luef, and P. H. Mayrhofer, Surf. Coat. Technol. (2011).

[46] Note:, $C_{ii}$s (i=1,2,3) ploted in Fig. 3a from the stress-strain method are dervied from the full elastic tensors and actually refer to the macroscopic cubic elastic constants.

[47] P. Jonnard, N. Capron, F. Semond, J. Massies, E. Martinez-Guerrero, and H. Mariette, EPJ B. **42**, 351 (2004).

[48] F. Litimein, B. Bouhafs, Z. Dridi, and P. Ruterana, New J. Phys. **4**, 64 (2002).

[49] K. Karch and F. Bechstedt, Phys. Rev. B **56**, 7404 (1997).

[50] D. Holec, R. Rachbauer, D. Kiener, P. D. Cherns, P. M. F. J. Costa, C. McAleese, P. H. Mayrhofer, and C. J. Humphreys, Phys. Rev. B **83**, 165122 (2011).

[51] B. D. Fulcher, X. Y. Cui, B. Delley, and C. Stampfl, Phys. Rev. B **85** (2012).

[52] A. Riedl, R. Daniel, M. Stefenelli, T. Schöberl, O. Kolednik, C. Mitterer, and J. Keckes, Scripta Mater., In press (2012).

[53] S. I. Ranganathan and M. Ostoja-Starzewski, Phys. Rev. Lett. **101** (2008).

[54] I. N. Frantsevich, F. F. Voronov, and S. A. Bokuta, Naukova Dumka, Kiev, 60 (1983).

[55] S. F. Pugh, Philos. Mag **45**, 823 (1954).

[56] D. G. Pettifor, Mater. Sci. Technol. **8**, 345 (1992).

Table I. Elastic tensors (in GPa) of c-Cr$_{1-x}$Al$_x$N calculated by stress-strain method.

| $x$ | $C_{11}$ | $C_{22}$ | $C_{33}$ | $C_{12}$ | $C_{13}$ | $C_{23}$ | $C_{44}$ | $C_{55}$ | $C_{66}$ | $C_{14}$ | $C_{15}$ | $C_{16}$ | $C_{24}$ | $C_{25}$ | $C_{26}$ | $C_{34}$ | $C_{35}$ | $C_{36}$ | $C_{45}$ | $C_{46}$ | $C_{56}$ |
|---|---|---|---|---|---|---|---|---|---|---|---|---|---|---|---|---|---|---|---|---|---|
| 1.00 | 426 | 426 | 426 | 167 | 167 | 167 | 306 | 306 | 306 | 0 | 0 | 0 | 0 | 0 | 0 | 0 | 0 | 0 | 0 | 0 | 0 |
| 0.89 | 433 | 431 | 433 | 161 | 160 | 160 | 274 | 274 | 274 | 0 | 1 | 0 | 0 | 1 | 0 | 0 | 1 | 0 | 0 | 0 | 0 |
| 0.78 | 442 | 455 | 453 | 150 | 148 | 146 | 246 | 246 | 247 | 5 | 0 | 0 | 8 | 0 | 0 | 9 | 0 | -1 | 0 | 0 | 1 |
| 0.67 | 461 | 466 | 464 | 143 | 145 | 142 | 220 | 224 | 220 | 0 | -3 | 2 | 2 | -3 | 1 | 0 | -2 | -1 | 0 | -1 | 1 |
| 0.56 | 466 | 494 | 468 | 134 | 140 | 131 | 188 | 202 | 193 | 2 | 3 | 5 | 2 | 2 | 4 | 1 | 0 | -1 | 0 | 1 | 0 |
| 0.44 | 474 | 504 | 488 | 129 | 136 | 123 | 167 | 183 | 175 | 1 | 1 | 4 | 5 | 0 | 1 | 4 | 0 | 0 | 1 | 1 | -1 |
| 0.33 | 493 | 510 | 492 | 117 | 123 | 124 | 164 | 163 | 159 | 1 | 2 | 15 | 3 | 2 | 9 | 1 | 0 | 8 | 1 | 0 | 2 |
| 0.22 | 513 | 513 | 505 | 117 | 118 | 119 | 150 | 141 | 141 | -4 | -4 | -3 | -4 | -1 | -4 | -3 | -3 | -2 | 0 | 0 | -1 |
| 0.11 | 515 | 518 | 499 | 119 | 117 | 115 | 123 | 132 | 123 | 1 | 2 | -3 | 1 | 0 | -5 | 1 | 2 | -4 | -1 | -1 | -1 |
| 0.00 | 521 | 515 | 512 | 111 | 118 | 115 | 109 | 127 | 111 | 3 | 9 | -9 | 5 | 0 | -8 | 8 | 8 | -6 | 1 | 0 | 0 |

**Figure caption**

FIG. 1. (Color online) Energy of formation, $E_f$, as a function of $x$ for $Cr_{1-x}Al_xN$ in the NM (solid square), PM (solid circle) states cubic B1 structures and PM state B4 structure (solid triangle).

FIG. 2. (Color online) (a) Lattice parameter, $a$, and (b) calculated bulk modulus, $B$, variation with $x$ for the NM (solid square) and PM (solid circle) state of c-$Cr_{1-x}Al_xN$. The inset of Fig. 2a shows the deviation from linear Vegard's-like behavior (bowing of the lattice parameter).

FIG. 3. (Color online) (a) Calculated elastic constants ($C_{11}$, $C_{12}$ and $C_{44}$), and (b) calculated bulk modulus, $B$, polycrystalline Young's modulus, $E$, and shear modulus, $G$ of PM c-$Cr_{1-x}Al_xN$. Solid squares and open triangle denote stress-strain and energy strain methods, respectively.

FIG. 4. (Color online) (a) Zener's anisotropy, $A$ and $E_{<111>}/E_{<100>}$ and (b) Poisson ratio $\nu$, $B/G$ and Cauchy pressure, $C_{12}$-$C_{44}$, of c-$Cr_{1-x}Al_xN$ in PM state. The values are based on stress-strain (solid square) and energy strain (solid triangle) methods.

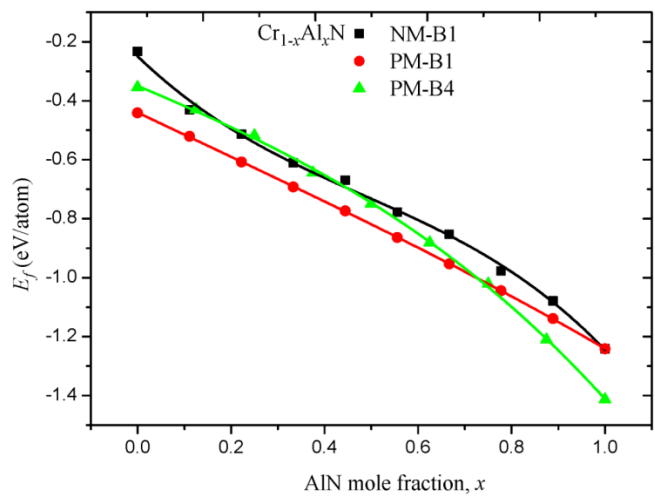

FIG. 1.

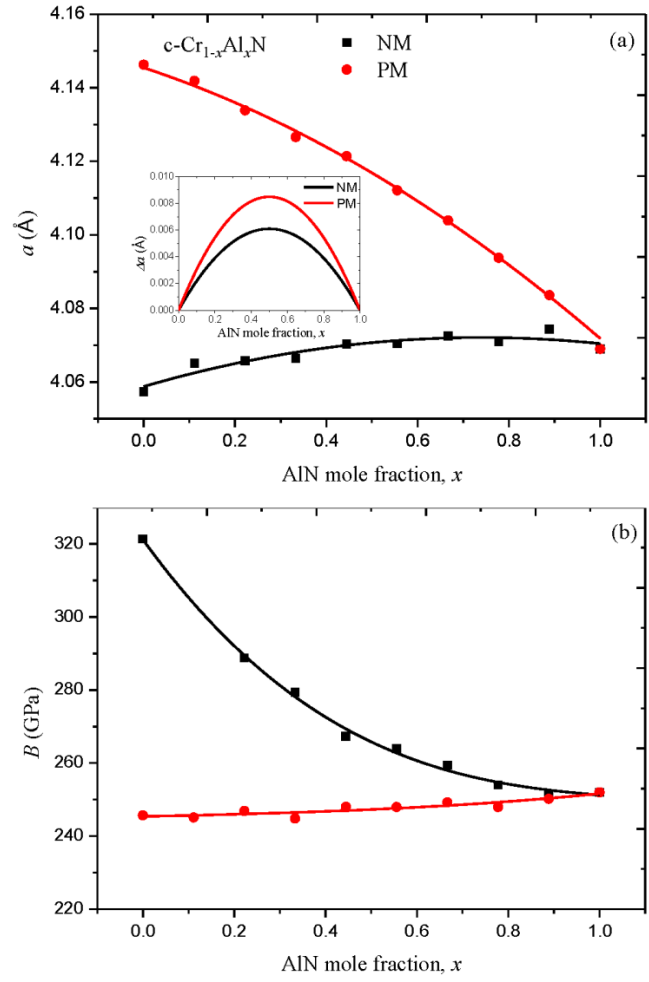

FIG. 2.

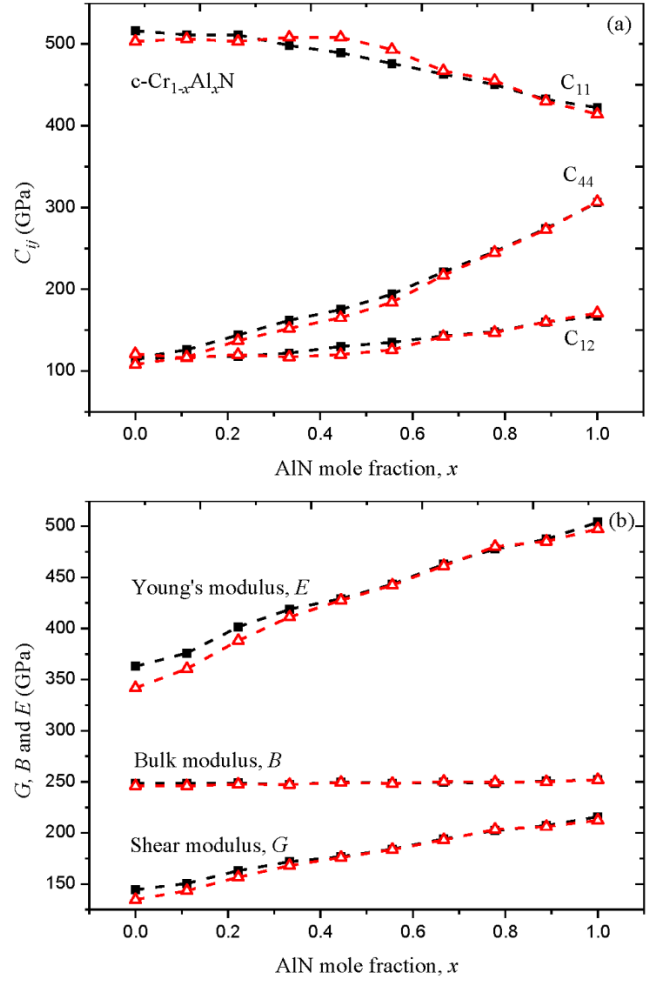

FIG. 3.

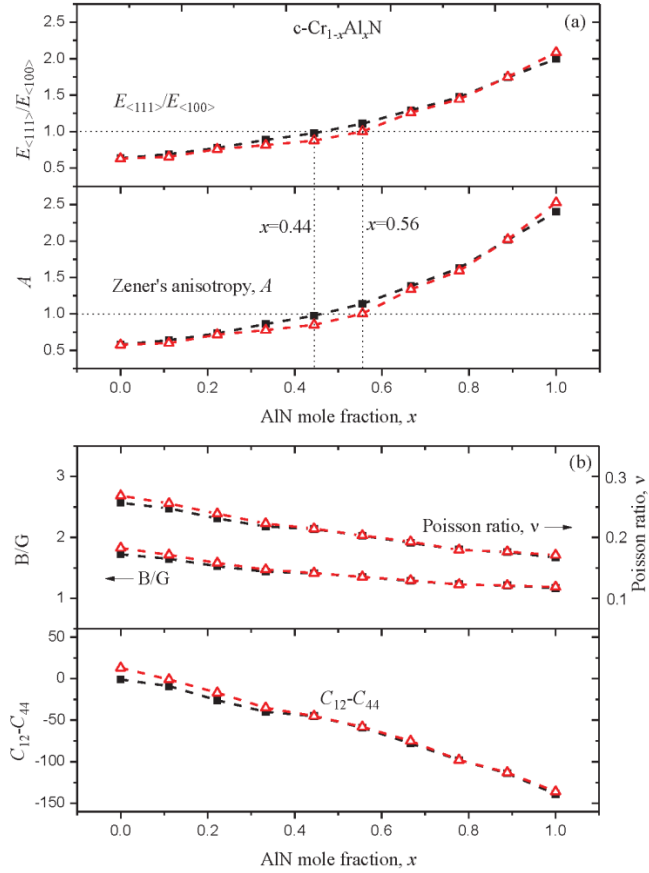

FIG. 4.

APPENDIX: SUPERCELL STRUCTURES AND SHORT RANGE ORDER PARAMETERS

TABLE II.

Arrangement of 3×3×2 supercells used for the calculation of the cubic B1 phases. The fractional coordinates are expressed in a cell with lattice vectors $a_1=a(1.5, 1.5, 0)$, $a_2=a(0, 1.5, 1.5)$, $a_3=a(1, 0, 1)$. Only atoms on the metallic sublattice are listed. The positions given here correspond to the starting configurations before any relaxation takes place.

| Fractional coordinate | | | Al content on the metallic sublattice | | | | | | | | |
|---|---|---|---|---|---|---|---|---|---|---|---|
| x | y | z | 0.000 | 0.111 | 0.222 | 0.333 | 0.444 | 0.556 | 0.667 | 0.778 | 0.889 |
| 0.667 | 0.000 | 0.000 | Cr↑ | Cr↓ | Cr↓ | Cr↑ | Cr↓ | Cr↓ | Al | Al | Cr↑ |
| 0.000 | 0.333 | 0.500 | Cr↑ | Cr↑ | Cr↑ | Cr↓ | Al | Al | Cr↑ | Al | Cr↓ |
| 0.000 | 0.000 | 0.000 | Cr↓ | Cr↑ | Cr↑ | Cr↓ | Cr↑ | Al | Cr↓ | Al | Al |
| 0.000 | 0.000 | 0.500 | Cr↓ | Cr↓ | Cr↓ | Cr↑ | Al | Al | Al | Cr↑ | Al |
| 0.000 | 0.333 | 0.000 | Cr↑ | Cr↑ | Cr↓ | Cr↓ | Cr↓ | Cr↑ | Al | Al | Al |
| 0.000 | 0.667 | 0.000 | Cr↑ | Cr↓ | Al | Al | Al | Al | Al | Cr↑ | Al |
| 0.000 | 0.667 | 0.500 | Cr↑ | Cr↑ | Cr↓ | Cr↓ | Cr↓ | Al | Al | Al | Al |
| 0.333 | 0.000 | 0.000 | Cr↑ | Cr↓ | Cr↑ | Al | Cr↑ | Al | Cr↑ | Cr↓ | Al |
| 0.333 | 0.000 | 0.500 | Cr↓ | Al | Cr↑ | Al | Cr↓ | Cr↑ | Al | Al | Al |
| 0.333 | 0.333 | 0.000 | Cr↓ | Cr↑ | Cr↑ | Al | Al | Cr↓ | Al | Al | Al |
| 0.333 | 0.333 | 0.500 | Cr↓ | Cr↓ | Cr↓ | Cr↑ | Al | Al | Al | Al | Al |
| 0.333 | 0.667 | 0.000 | Cr↓ | Al | Cr↓ | Cr↑ | Al | Al | Cr↓ | Al | Al |
| 0.333 | 0.667 | 0.500 | Cr↓ | Cr↓ | Al | Cr↑ | Cr↑ | Cr↓ | Cr↑ | Al | Al |
| 0.667 | 0.000 | 0.500 | Cr↑ | Cr↑ | Cr↑ | Cr↓ | Cr↓ | Cr↓ | Cr↓ | Al | Al |
| 0.667 | 0.333 | 0.000 | Cr↓ | Cr↓ | Al | Cr↑ | Al | Al | Al | Al | Al |
| 0.667 | 0.333 | 0.500 | Cr↑ | Cr↓ | Al | Al | Cr↑ | Cr↑ | Al | Al | Al |
| 0.667 | 0.667 | 0.000 | Cr↑ | Cr↑ | Cr↓ | Cr↓ | Cr↑ | Cr↑ | Al | Al | Al |
| 0.667 | 0.667 | 0.500 | Cr↓ | Cr↑ | Cr↑ | Al | Al | Al | Al | Cr↓ | Al |

Table III.

Arrangement of 2×2×2 supercells used for the calculation of the wurtzite B4 phases. The fractional coordinates are expressed in a cell with lattice vectors $a_1=a(1, 0, 0)$, $a_2= a (-0.5, 0.866, 0)$, $a_3= a (0, 0, c/a)$. Only atoms on the metallic sublattice are listed. The positions given here correspond to the starting configurations before any relaxation takes place.

| Fractional coordinate | | | Al content on the metallic sublattice | | | | | | | |
|---|---|---|---|---|---|---|---|---|---|---|
| x | y | z | 0.000 | 0.125 | 0.250 | 0.375 | 0.500 | 0.625 | 0.750 | 0.875 |
| 0.167 | 0.333 | 0.500 | Cr↑ | Cr↓ | Cr↓ | Al | Al | Cr↓ | Al | Cr↑ |
| 0.667 | 0.333 | 0.500 | Cr↑ | Cr↑ | Cr↓ | Cr↓ | Al | Cr↑ | Al | Cr↓ |
| 0.167 | 0.333 | 0.000 | Cr↓ | Cr↓ | Cr↑ | Cr↑ | Al | Al | Al | Al |
| 0.333 | 0.167 | 0.250 | Cr↑ | Cr↓ | Cr↑ | Cr↓ | Cr↑ | Al | Al | Al |
| 0.333 | 0.167 | 0.750 | Cr↓ | Cr↓ | Cr↓ | Al | Al | Al | Al | Al |
| 0.167 | 0.833 | 0.000 | Cr↓ | Cr↑ | Cr↑ | Cr↑ | Cr↓ | Al | Cr↓ | Al |
| 0.333 | 0.667 | 0.250 | Cr↑ | Cr↑ | Cr↓ | Al | Al | Cr↓ | Al | Al |
| 0.167 | 0.833 | 0.500 | Cr↑ | Cr↑ | Cr↓ | Al | Cr↓ | Cr↑ | Al | Al |
| 0.333 | 0.667 | 0.750 | Cr↓ | Cr↓ | Cr↓ | Al | Al | Cr↓ | Al | Al |
| 0.667 | 0.333 | 0.000 | Cr↑ | Cr↓ | Al | Cr↓ | Cr↑ | Al | Al | Al |
| 0.833 | 0.167 | 0.250 | Cr↓ | Cr↑ | Al | Cr↓ | Cr↓ | Al | Al | Al |
| 0.833 | 0.167 | 0.750 | Cr↓ | Al | Al | Cr↑ | Al | Al | Cr↑ | Al |
| 0.667 | 0.833 | 0.000 | Cr↓ | Cr↑ | Cr↑ | Cr↓ | Cr↑ | Al | Al | Al |
| 0.833 | 0.667 | 0.250 | Cr↑ | Cr↑ | Cr↑ | Cr↑ | Cr↑ | Al | Cr↓ | Al |
| 0.667 | 0.833 | 0.500 | Cr↓ | Al | Cr↑ | Al | Al | Al | Al | Al |
| 0.833 | 0.667 | 0.750 | Cr↑ | Cr↓ | Al | Cr↑ | Cr↓ | Cr↑ | Cr↑ | Al |

Table IV.

Short-range order (SRO) parameters for the supercells used to model the different fractions of AlN in this work (Tables II and III). The shells correspond to nearest distances between atoms on the metallic sublattice.

| | Cubic B1 | | | | | | Wurtzite B4 | | | | |
| --- | --- | --- | --- | --- | --- | --- | --- | --- | --- | --- | --- |
| | | | Shell | | | | | | Shell | | |
| AlN% | 1 | 2 | 3 | 4 | 5 | AlN% | 1 | 2 | 3 | 4 | 5 |
| 0.889 | -0.09 | 0.00 | -0.05 | 0.09 | -0.05 | 0.875 | -0.01 | -0.10 | -0.10 | -0.04 | -0.10 |
| 0.778 | -0.10 | 0.00 | 0.00 | 0.00 | -0.05 | 0.750 | -0.05 | -0.08 | 0.08 | -0.06 | -0.13 |
| 0.667 | -0.11 | 0.00 | 0.00 | 0.00 | -0.02 | 0.625 | 0.00 | -0.04 | 0.07 | 0.00 | -0.31 |
| 0.556 | -0.09 | 0.00 | -0.02 | -0.02 | 0.00 | 0.500 | -0.03 | 0.00 | 0.00 | -0.02 | -0.27 |
| 0.444 | -0.10 | 0.00 | -0.02 | 0.00 | 0.00 | 0.375 | 0.00 | -0.04 | 0.06 | 0.00 | -0.32 |
| 0.333 | -0.08 | 0.00 | -0.04 | 0.00 | 0.00 | 0.250 | 0.00 | -0.05 | 0.05 | -0.02 | -0.27 |
| 0.222 | -0.10 | 0.00 | -0.02 | 0.00 | 0.00 | 0.125 | -0.02 | -0.04 | -0.04 | -0.02 | -0.25 |
| 0.111 | -0.09 | 0.00 | -0.02 | -0.02 | 0.00 | 0.000 | 0.00 | 0.00 | 0.00 | 0.00 | -0.33 |
| 0.000 | -0.04 | -0.04 | -0.09 | 0.04 | 0.00 | | | | | | |